\documentclass[aps, prl, nobibnotes, twocolumn, superscriptaddress, bibliography]{revtex4-2}
\usepackage{mathtools}
\usepackage{amsfonts}
\usepackage{mathrsfs}
\usepackage{amsmath}
\usepackage{color}
\usepackage{graphicx}
\usepackage{bm}
\usepackage{amssymb}
\usepackage{xspace}
\usepackage{epstopdf}
\usepackage{dcolumn}
\usepackage{longtable}
\usepackage{multirow}
\usepackage{float}
\usepackage{comment}
\usepackage{pifont}
\usepackage{bbding}
\usepackage{fontawesome}
\usepackage{soul}
\usepackage{esint}
\usepackage{amsmath}
\usepackage{braket}

\usepackage[colorlinks=true,  letterpaper=true,  pdfstartview=FitV,  linkcolor=blue,  citecolor=blue,  urlcolor=blue]{hyperref}

\makeatother

\begin{document}
\title{Equivariant  Space Group and  Hamiltonian for  Collinear Magnetic Systems}
\author{Chaoxi Cui}
\affiliation{Centre for Quantum Physics, Key Laboratory of Advanced Optoelectronic
Quantum Architecture and Measurement (MOE), School of Physics, Beijing
Institute of Technology, Beijing 100081, China}
\affiliation{Beijing Key Lab of Nanophotonics \& Ultrafine Optoelectronic Systems,
School of Physics, Beijing Institute of Technology, Beijing 100081,
China}

\author{Zhi-Ming Yu}
\email{zhiming\_yu@bit.edu.cn}
\affiliation{Centre for Quantum Physics, Key Laboratory of Advanced Optoelectronic
Quantum Architecture and Measurement (MOE), School of Physics, Beijing
Institute of Technology, Beijing 100081, China}
\affiliation{Beijing Key Lab of Nanophotonics \& Ultrafine Optoelectronic Systems,
School of Physics, Beijing Institute of Technology, Beijing 100081,
China}

\author{Yilin Han}
\affiliation{Centre for Quantum Physics, Key Laboratory of Advanced Optoelectronic
Quantum Architecture and Measurement (MOE), School of Physics, Beijing
Institute of Technology, Beijing 100081, China}
\affiliation{Beijing Key Lab of Nanophotonics \& Ultrafine Optoelectronic Systems,
School of Physics, Beijing Institute of Technology, Beijing 100081,
China}

\author{Run-Wu Zhang}
\affiliation{Centre for Quantum Physics, Key Laboratory of Advanced Optoelectronic
Quantum Architecture and Measurement (MOE), School of Physics, Beijing
Institute of Technology, Beijing 100081, China}
\affiliation{Beijing Key Lab of Nanophotonics \& Ultrafine Optoelectronic Systems,
School of Physics, Beijing Institute of Technology, Beijing 100081,
China}

\author{Shengyuan A. Yang}
\email{shengyuan.yang@polyu.edu.hk}
\affiliation{Research Laboratory for Quantum Materials, Department of Applied Physics,
	The Hong Kong Polytechnic University, Kowloon, Hong Kong, China}

\author{Yugui Yao}
\email{ygyao@bit.edu.cn}
\affiliation{Centre for Quantum Physics, Key Laboratory of Advanced Optoelectronic
Quantum Architecture and Measurement (MOE), School of Physics, Beijing
Institute of Technology, Beijing 100081, China}
\affiliation{Beijing Key Lab of Nanophotonics \& Ultrafine Optoelectronic Systems,
School of Physics, Beijing Institute of Technology, Beijing 100081,
China}
\affiliation{Beijing Institute of Technology, Zhuhai 519000, China}

\begin{abstract}
Condensed matter physics increasingly focuses on exploiting the magnetic order parameter orientation \(\hat{\bm n}\)  as a tuning knob for properties of collinear magnetic  materials, but a general method for constructing effective Hamiltonians with explicit \(\hat{\bm n}\)-dependence has been lacking.
Here, we develop a symmetry-based framework, built on the equivariant space group, for constructing such Hamiltonians, termed equivariant magnetic Hamiltonians (EMHs).
The resulting EMH lives in a higher-dimensional \(\bm k\)-\(\hat{\bm n}\) space and exhibits unconventional symmetry actions and topological features.
Using a 1D ferromagnetic chain and a 3D antiferromagnet as examples, we demonstrate that explicit \(\hat{\bm n}\)-dependence in EMHs enables the study of magnetic-dynamics-driven topological pumping, including even-integer charge pumping and a second-Chern-number-induced  quantized pumping of surface anomalous Hall conductivity.
Beyond model systems, we incorporate the framework into first-principles calculations to construct \emph{ab-initio} EMHs that accurately capture the \(\hat{\bm n}\)-dependent band structures of real materials. The approach can also be generalized to non-collinear magnetic systems.
Our work establishes a general framework for constructing EMHs and for exploring the rich physics arising from magnetic anisotropy and magnetic dynamics.

\end{abstract}
\maketitle

Magnetic materials have been a major topic in condensed matter physics research. The spontaneous breaking of time-reversal $\mathcal T$ symmetry in these systems {is} described by the formation of magnetic order parameter, e.g., magnetization vector $\bm M$ for ferromagnets and N\'{e}el vector $\bm N$ for collinear antiferromagnets, which are defined from the configuration of local magnetic moment. For temperatures well below the transition temperature, the magnitude of the order parameter vector is usually frozen, remaining largely unchanged under perturbations~\cite{Blundell2001Magnetism,Auerbach1994QuantumMagnetism}. Meanwhile, the orientation of $\bm M$ or $\bm N$, denoted by a unit vector $\hat{\bm n}$, is a degree of freedom that can be tuned. In fact, a central goal of spintronics is to effectively switch the direction $\hat{\bm n}$, which constitutes the basis of information storage technology and, by now, can be achieved via various methods~\cite{RevModPhys.76.323,RevModPhys.77.1375,Chappert2007NatMaterDataStorage,RALPH20081190,RevModPhys.90.015005,RevModPhys.91.035004,RevModPhys.82.2731,RevModPhys.96.015005,SongCheng2025Nat,SongCheng2026Nat,MnTe2024Nat,SciAdv2024}. On the other hand, many properties and effects of magnets, such as anomalous Hall effect, magneto-optical effects, and magnetoresistance, exhibit strong dependence on $\hat{\bm n}$~\cite{
HIRSCH1973239,PhysRevB.77.024409,PhysRevB.78.214435,  Koizumi2023NatCommunQuadrupoleAHE,Sankar2026NatCommunIPHE,Liu2025PRXMultipolarAHE,Xiao2026SCPMAAHNT,
MagnetoOpticalEllipsometer, DAAllwood_2003,	Pan2026PRLNeelMOKE,
fina2014anisotropic,Yang2018NatCommunAHMR,Kriegner2024npjSpintronicsMnTeAMR
}. To explore the control of $\hat{\bm n}$ as a tuning knob of magnetic materials' properties is a hot topic of current research.

In studying physical properties of materials, effective Hamiltonians constructed with symmetry constraints have been a powerful tool~\cite{roland2003spin,YU2022375}.
They not only form starting points for almost all theoretical model studies, but are also widely used (e.g., the \emph{ab-initio} tight-binding Hamiltonians) in first-principles approaches~\cite{Mostofi2008Wannier90,Marzari2012RMPWannier}. For magnetic systems, to investigate consequences of the varying direction of magnetic order parameter, it naturally requires an effective Hamiltonian $\mathcal H(\hat{\bm n})$ that \emph{explicitly exhibits its dependence on} $\hat{\bm n}$. Unfortunately, a systematic approach to
construct such Hamiltonians has not been developed yet.

In this work, we solve this outstanding problem. Focusing on collinear magnetic systems, we show that $\mathcal H(\hat{\bm n})$, referred to as an equivariant magnetic Hamiltonian (EMH), is constructed by constraints of an equivariant space group (ESG),  different from conventional magnetic/spin space groups that are commonly adopted for describing magnetic systems. The resulting EMH  lives on a higher-dimensional $\bm k$-$\hat{\bm n}$ space with unusual symmetry actions and interesting topological consequences. As examples, we demonstrate EMHs constructed for a 1D ferromagnetic chain and for a 3D antiferromagnet.
The explicit $\hat{\bm n}$-dependence in EMH enables the study of topological pumping driven by magnetic dynamics. For the 1D chain, the rotation of magnetization vector realizes a charge pump. Interestingly, the number $q$ of electrons pumped per cycle must be an even integer, as required by ``time-reversal'' symmetry. And certain crystalline symmetry may reduce the fundamental domain of $\bm k$-$\hat{\bm n}$ space, putting further constraints on $q$, e.g., a $p$-fold rotation would require $q\in p\mathbb Z$. Meanwhile, the 3D antiferromagnetic EMH may realize a pump of quantized surface anomalous Hall conductivity
characterized by a second Chern number. Beyond model study, we demonstrate incorporation of our approach with first-principles calculations to construct an \emph{ab-initio} EMH, which offers an accurate description of $\hat{\bm n}$ dependence for real materials, as evidenced in the example of monolayer MnBi$_2$Te$_4$. Furthermore, we discuss how our approach can be extended to general non-collinear magnetic systems. Our work provides a general solution for constructing EMHs, which establishes the basis for exploring the rich physics induced by anisotropy associated with magnetic orders and by magnetic dynamics.

\textit{\textcolor{blue}{Equivariant space group.}}
Consider a collinear magnetic system, which can have either ferromagnetic or antiferromagnetic (including altermagnetic) ordering~\cite{Blundell2001Magnetism,PhysRevX.12.040501}.
For simplicity, assume the ferromagnetic (antiferromagnetic) case has only one (two) magnetic sublattice(s).
The extension to multiple sublattices is straightforward. The magnetic order parameter for collinear magnets is a pseudovector. Our task is to construct the EMH with explicit dependence on the direction $\hat{\bm n}$ of this pseudovector.

\begin{figure}
	\includegraphics[width=8. cm]{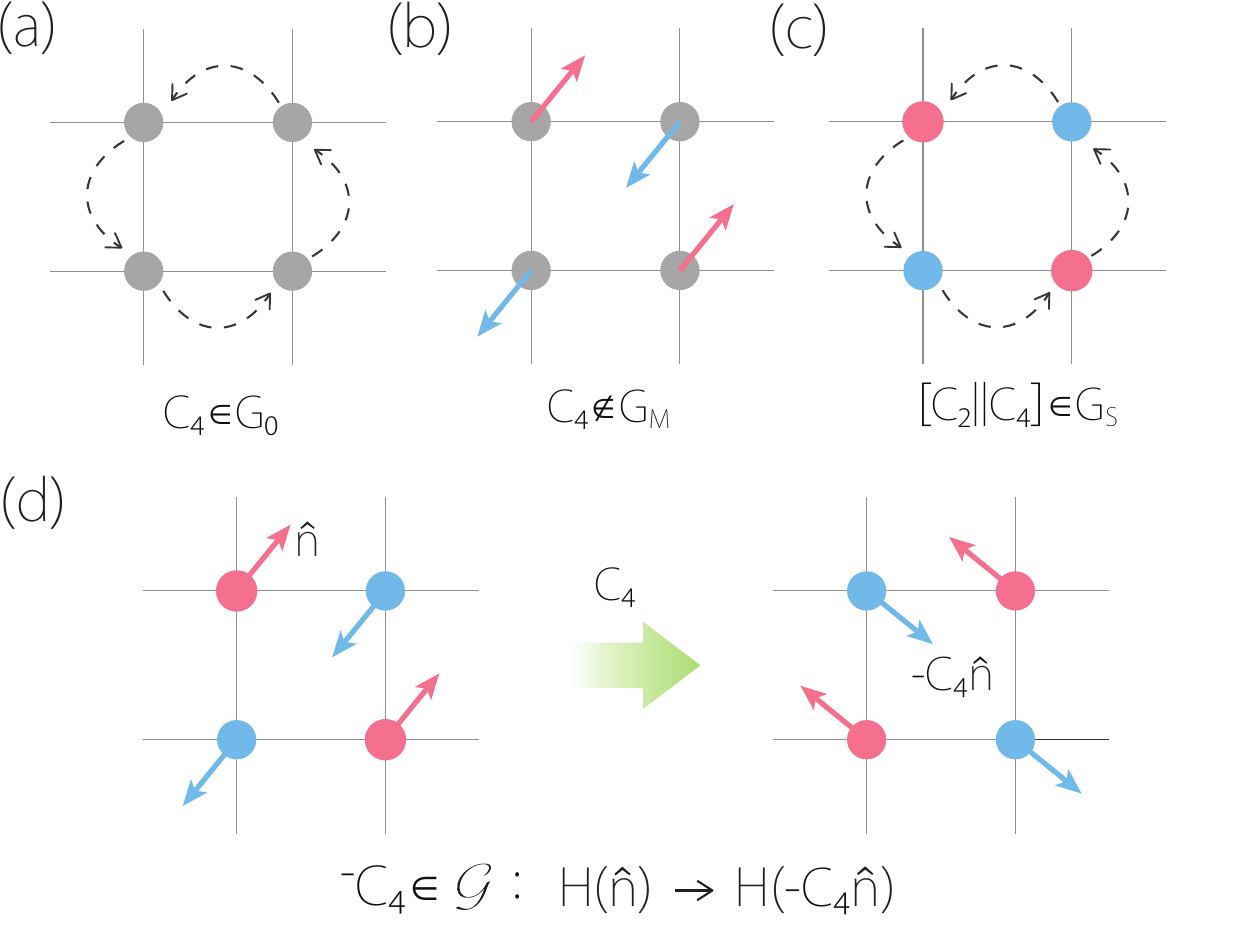}
\caption{ A simple square lattice.  (a) Without magnetism, its nonmagnetic space group $G_0$ contains $C_4$ symmetry.
	 (b) With an in-plane antiferromagnetic order, $C_4$ symmetry is broken, so $C_4$ is not in the magnetic space group $G_\mathrm{M}$.
	 (c) For spin space group $G_\mathrm{S}$, the magnetic order manifests as different colors of the two sublattices, whereas the information of direction $\hat{\bm n}$ is lost. (d) The ESG element $\prescript{-}{}C_4$ connects two antiferromagnetic configurations on the same lattice. Here, $\zeta=-$ indicates that the $C_4$ rotation switches the two sublattices. This ESG element is required for constraining the EMH.
	}
	\label{Fig.1}
\end{figure}
Let's first recall how conventional effective Hamiltonian is constructed. One usually starts with a lattice and a set of active local orbitals. For purely model studies, these are simply chosen to satisfy the imposed symmetry; for modeling real materials, the choice is made according to the material's structure and the energy bands one aims to describe.
From these local orbitals, one then obtains a set of Bloch-like basis, over which the effective Hamiltonian $H(\bm k)$ is constructed in $k$-space. The key step of the construction is to go through the symmetry constraint relations for fixing all possible terms in $H$ and their interrelations. Such relations have a generic form~\cite{MagneticKP}:
\begin{equation}\label{HH}
  D(O)H({\bm k})D(O)^{-1} = H(O{\bm k}),
\end{equation}
where $O$ is an element of the governing symmetry group of the system, and $D(O)$ denotes its representation in the basis of chosen orbitals.

A magnetic system is commonly described by two kinds of symmetry groups: magnetic space group $G_\text{M}$ and spin space group $G_\text{S}$, depending on whether spin-orbit coupling (SOC) is considered or not~\cite{BradleyCracknell2009SymmetrySolids,Litvin2013MagneticGroupTables,1966RSPSA.294..343B,LITVIN1974538,Xiao2024PRXSSG,Chen2024PRXSSG,Jiang2024PRXSSG,LiuQihangNature}. With SOC, $O$ in Eq.~(\ref{HH}) should go through $G_\text{M}$. However, the group $G_\text{M}$ (hence the resulting $H$) is tied to a specific $\hat{\bm n}$; and $G_\text{M}$ in general varies with the direction $\hat{\bm n}$ [see Fig.~\ref{Fig.1}(a,b)], so it does not allow us to obtain a single Hamiltonian hosting $\hat{\bm n}$-dependence. On the other hand, with $O\in G_\text{S}$, one may construct an $H$ in the absence of SOC. However, since spin and spatial degrees of freedom are completely decoupled in this case  [see Fig.~\ref{Fig.1}(c)], the obtained $H$ will not have any $\hat{\bm n}$-dependence.

To acquire explicit $\hat{\bm n}$-dependence, it is essential to place $\hat{\bm n}$ on equal footing with $\bm k$ in a Hamiltonian $\mathcal H(\bm k, \hat{\bm n})$, which lives in the $\bm k$-$\hat{\bm n}$ space.
And the symmetry operations $O$ should act on both $\bm k$ and $\hat{\bm n}$. As discussed, the governing symmetry group here is not $G_\text{M}$ nor $G_\text{S}$ (which leave the system invariant), but some group $\mathcal G$ that may rotate $\hat{\bm n}$, establishing connection between different points of $\bm k$-$\hat{\bm n}$ space. In this sense, we name  group $\mathcal G$ as an ESG.

The ESG $\mathcal G$ is constructed as follows. It consists of elements of the form $^{\zeta}X$. Here, $X$ belongs to the nonmagnetic space group $G_0$ (i.e., when one neglects the magnetic ordering on the lattice), while $\zeta\in \mathbb Z_2$ is \emph{fixed} by $X$: $\zeta=+$ if $X$ preserves each magnetic sublattice, and $\zeta=-$ if $X$ switches two sublattices. Note that time-reversal operation $\mathcal T$ does not act on lattice, so $\mathcal G$ always contains a `time-reversal' symmetry: $^+\mathcal T$. The product rule of the group is simply
\begin{equation}
\prescript{\zeta_1}{}X_1 \cdot \prescript{\zeta_2}{}X_2 =\prescript{(\zeta_1\cdot \zeta_2)}{}(X_1\cdot X_2).
\end{equation}
As an abstract group, $\mathcal G$ is isomorphic to $G_0$. However, $\mathcal G$ contains also information of magnetic order (via $\zeta$), which is not in $G_0$.

The necessity to adopt ESG for constraining EMH $\cal H$ can be readily understood from Fig.~\ref{Fig.1}(d). It shows two different antiferromagnetic states on a square lattice, yet they are connected by a fourfold rotation $C_4$. Note that (i) $C_4$ is not a symmetry of the magnetic lattice, instead, it is a symmetry for the nonmagnetic lattice; (ii)
$C_4$ switches the two magnetic sublattices, so the N\'{e}el unit vector  {of the right configuration in Fig.~\ref{Fig.1}(d)} should be $-C_4 \hat{\bm n}$ if  {denoting that of the left configuration by $\hat{\bm n}$}, and this switching character must be captured in the symmetry element. This discussion demonstrates that $\mathcal H(\hat{\bm n})$ and $\mathcal H(-C_4\hat{\bm n})$ are not independent, and their connection can be properly described by an element $\prescript{-}{}C_4$ belonging to ESG.

Based on the observation, one naturally finds the following constraint relations: For any $\prescript{\zeta}{}X\in \cal G$,
\begin{eqnarray}\label{SymFM}
	D(\prescript{\zeta}{}X){\cal H}({\bm k}, \hat{\boldsymbol{n}})D(\prescript{\zeta}{}X)^{-1} = {\cal H}(X{\bm k}, \zeta X \hat{\boldsymbol{n}}).
\end{eqnarray}
This captures the fact that points $({\bm k}, \hat{\boldsymbol{n}})$ and $(X{\bm k}, \zeta X \hat{\boldsymbol{n}})$ in $\bm k$-$\hat{\bm n}$ space are connected by the $\prescript{\zeta}{}X$ symmetry operation. The
inclusion of $\zeta$ in $\zeta X \hat{\boldsymbol{n}}$ is crucial, for correctly capturing the possible sublattice switch.
And $D(\prescript{\zeta}{}X)$ can be directly obtained from the representation matrix of $X$ for
the nonmagnetic lattice. Using the constraints (\ref{SymFM}) by ESG, the required EMH can be constructed.

\textit{\textcolor{blue}{Symmetry action and form factor matrix.}}
EMH lives on the $\bm k$-$\hat{\bm n}$ space. It has the topology of $T^d\times S^2$ ($d$ is spatial dimension of the system), which is no longer a torus. The action of symmetry operation on $\hat{\bm n}$ also exhibit interesting features distinct from that on $k$ space. For example, since $\hat{\bm n}$ is an axial vector, it is invariant under inversion $\prescript{+}{}P$. If we parameterize $\hat{\bm n}$ with spherical angles $(\theta_{\hat{\bm n}},\phi_{\hat{\bm n}})$, the action of $p$-fold rotation
$
\prescript{+}{}C_{pz}: (\theta_{\hat{\bm n}},\phi_{\hat{\bm n}})\mapsto (\theta_{\hat{\bm n}}, \phi_{\hat{\bm n}}+2\pi/p)
$, is like a translation in {$\phi_{\hat{\bm n}}$}. Meanwhile, the action
$\prescript{+}{}{\cal T}: (\theta_{\hat{\bm n}},\phi_{\hat{\bm n}})\mapsto (\pi-\theta_{\hat{\bm n}}, \phi_{\hat{\bm n}}+\pi)$
mimics a glide mirror. In the example below, we shall see the important consequences of such nontrivial actions.

%
%

It is convenient to describe the $\hat{\bm n}$-dependence of EMH with an expansion in spherical harmonics:
\begin{eqnarray}
	{\cal H}({\bm k}, \hat{\boldsymbol{n}}) = \sum_{l=0}^{\infty}\sum_{m=1}^{2l+1} {\cal{U}}_{lm}(\boldsymbol{k}) Y_{lm}(\theta_{\hat{\bm n}},\phi_{\hat{\bm n}}) \label{HamExpand},
\end{eqnarray}
where $Y_{lm}$ are \emph{real} spherical harmonics, and each ${\cal{U}}_{lm}$ is a hermitian matrix. Although $l$ has no upper bound, in practice, the terms quickly decay with $l$, so it is usually sufficient to retain the first few terms.

Performing this expansion in Eq.~(\ref{SymFM}), one obtains the following constraints on ${\cal{U}}_{lm}$:
\begin{eqnarray}\label{HknConstrain}
	D(\prescript{\zeta}{}X) {\cal{U}}_{lm}D(\prescript{\zeta}{}X)^{-1}=\zeta^{l}\sum_{n=1}^{2l+1} {\cal{U}}_{ln}(X \boldsymbol{k}) [ S_{l}(X)]_{nm},
\end{eqnarray}
where
\begin{eqnarray}\label{FFM}
	[ S_{l}(X)]_{nm}=\oint Y_{ln}(X\hat{\bm n})Y_{lm}(\hat{\bm n})d\Omega_{\hat{\bm n}},
\end{eqnarray}
is a form factor matrix depending only on the rotational part of $X$, and $\Omega_{\hat{\bm n}}$ is the solid angle. The $S_l$ matrices for typical point group elements are tabulated in Supplemental Material (SM)~\cite{SM}. \nocite{kresse1996efficient,kresse1996efficiency,blochl1994projector,kresse1999ultrasoft,perdew1996gga,perdew1998reply,monkhorst1976special,dudarev1998electron,marzari1997maximally}
And Eq.~(\ref{HknConstrain}) can be readily implemented to obtain the EMH.

\begin{figure}
	\includegraphics[width=8 cm]{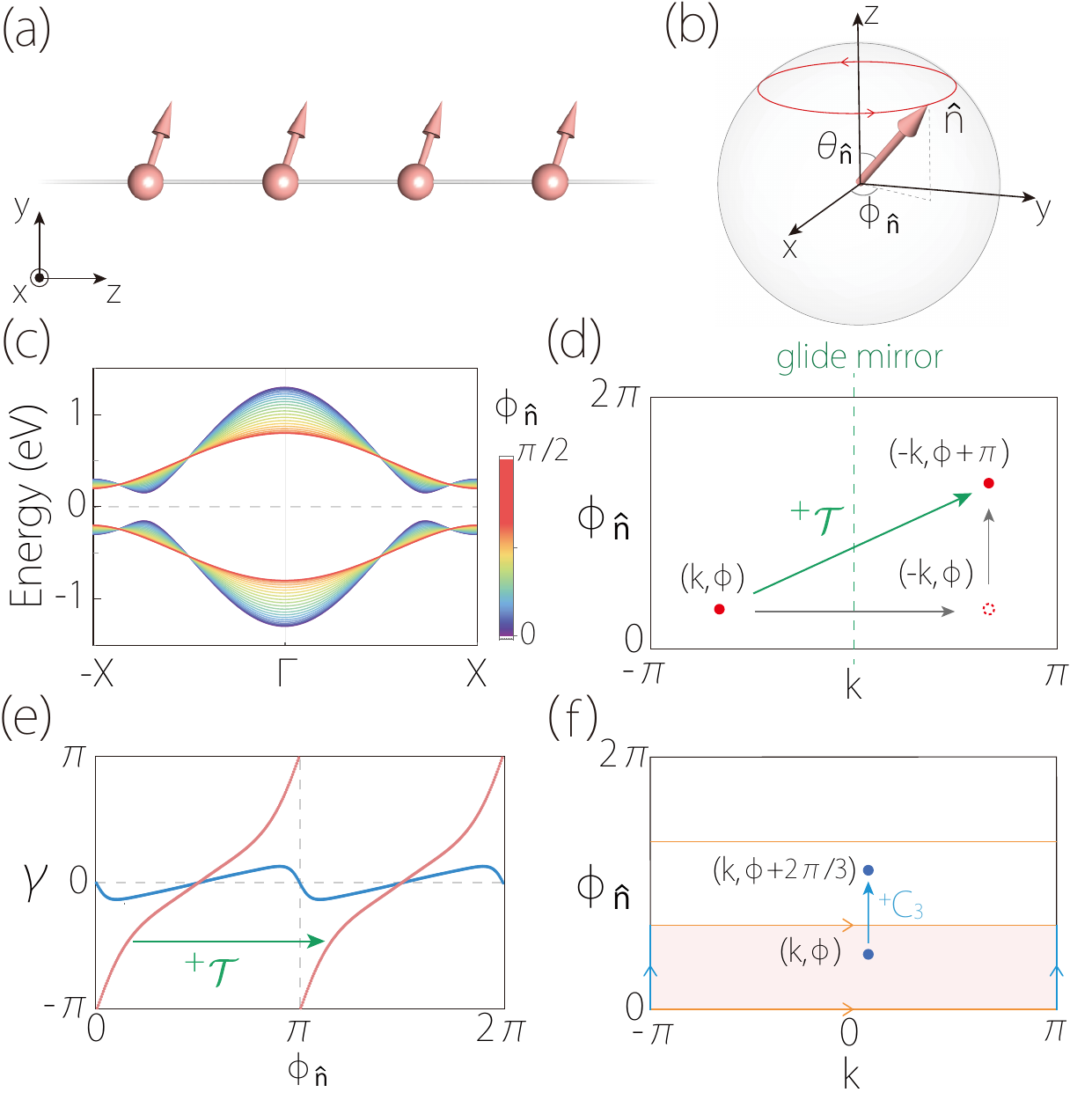}
	\caption{
(a) Schematic of 1D ferromagnetic chain. (b) Precession of $\hat{\bm n}$. (c) Band structures of  EMH (\ref{H2D}) for different values of $\phi_{\hat{\bm n}}$ with $\theta_{\hat{\bm n}}=\pi/2$. The system remains an insulator in this case. (d) The action of $\prescript{+}{}{\cal T}$ on the $k$-$\phi_{\hat{\bm n}}$ space, which resembles a glide mirror. (e) Evolution of the Zak phase $\gamma(\phi_{\hat{\bm n}})$. $\prescript{+}{}{\cal T}$ symmetry requires $\gamma$ having a $\pi$ periodicity in $\phi_{\hat{\bm n}}$.
The red and the blue lines are the cases giving $q=2$ (red) and $q=0$ (blue), respectively.
(f) Rotational symmetry like $\prescript{+}{}C_3$ acts like a fractional translation on $k$-$\phi_{\hat{\bm n}}$ space. Here, the fundamental domain of the space (highlighted in red color) is $1/3$ of the original space.
For calculations in (c) and (e), we take $t_0=0$, $s_0=0.2$~eV, $M_1=M_2=0.5$~eV; $t_1=0.8$~eV and $t_2=0.3$~eV for (c) and the red line in (e), whereas $t_1=0.45$~eV and $t_2=0.2$~eV for the blue line in (e).
	}
	\label{Fig.2}
\end{figure}

\textit{\textcolor{blue}{1D ferromagnetic chain \& $2\mathbb Z$ charge pump.}}
Let's first apply our approach to construct EMH for a simple 1D ferromagnetic chain. As illustrated in Fig.~\ref{Fig.2}(a), assume the chain has one active site per unit cell, on which there are two $s$-like spin-polarized orbitals $\ket{\uparrow}$ and $\ket{\downarrow}$.
Take $G_0=T_z\rtimes D_2$, where $T_z$ is the 1D translation group along $z$ and the point group $D_2$ consists of three
orthogonal two-fold rotations~\cite{SM}. For this ferromagnet, the $\zeta$ index is always $+$. Then, the generator elements of ESG $\cal G$ that constrain EMH can be chosen as $\prescript{+}{}C_{2x}$, $\prescript{+}{}C_{2z}$, and $\prescript{+}{}{\cal{T}}$~\footnote{Here, the translation part has trivial action on ${\cal H}({\bm k}, \hat{\boldsymbol{n}})$.}, with the following representation in the given orbital basis: $D(\prescript{+}{}C_{2x})=-i\sigma_1$, $D(\prescript{+}{}C_{2z})=-i \sigma_3$, and $D(\prescript{+}{}{\cal T})=-i\sigma_2 K$, where $\sigma$'s are Pauli matrices and $K$ is complex conjugation.

Following Eq.~(\ref{HamExpand}), let's construct an EMH up to $l=1$, which corresponds to
\begin{eqnarray}\label{77}
	{\cal H} = {\cal{U}}_{01}Y_{01}  +\sum_{j=1}^3 {\cal U}_{1j}Y_{1j}(\theta_{\hat{\bm n}},\phi_{\hat{\bm n}}).
\end{eqnarray}
Here, the $\hat{\bm n}$-dependence is contained in the three real spherical harmonics $Y_{1j}$, and the remaining task is to
find the $k$-dependent $\cal U$ matrices, based on Eq.~(\ref{HknConstrain}). The computation is straightforward (see SM~\cite{SM}). Up to nearest neighboring hopping, we obtain
\begin{eqnarray}
	{\cal{U}}_{01} &= & t_{0} \cos k+ s_{0} \sin k\, \sigma_{3}, \\
{\cal{U}}_{1j} &=&  (M_j + t_{j} \cos k   ) \sigma_{j}+\delta_{j,3} s_{1} \sin k,
\end{eqnarray}
where $t$'s and $s$'s are real parameters. Note that in ${\cal{U}}_{1j}$, if putting $t_j=s_1=0$, then the $\hat{\bm n}$-dependence would be reduced to the familiar Zeeman coupling $\propto \hat{\bm n}\cdot\bm \sigma$. Hence, the other terms represent the nontrivial $\hat{\bm n}$-dependent effects that are captured by EMH.

The EMH allows us to investigate how the magnetization direction affects electronic properties. Particularly, we are interested in possible adiabatic charge pumping by magnetic precession [see Fig.~\ref{Fig.2}(b)]. For simplicity, assume the magnetization has an easy-plane anisotropy, such that the dynamics of $\hat{\bm n}$ is confined in the $x$-$y$ plane, i.e., $\theta_{\hat{\bm n}}=\pi/2$. In this case, $\cal H$ is defined on the 2D $k$-$\phi_{\hat{\bm n}}$ space (which has topology of a torus $T^2$), with the following form:
\begin{eqnarray}\label{H2D}
{\cal H}(k,\phi_{\hat{\bm n}}) &= & t_0\cos k+(M_1+t_1\cos k)\cos\phi_{\hat{\bm n}}\sigma_1\nonumber\\
 &+&(M_2+t_2\cos k)\sin\phi_{\hat{\bm n}}\sigma_2+s_0 \sin k\,\sigma_3,
\end{eqnarray}
where all normalization factors have been absorbed into the model parameters for simplicity. Figure~\ref{Fig.2}(c) shows evolution of the band structure as $\phi_{\hat{\bm n}}$ varies.

Assume the system remains an insulator during the evolution of $\phi_{\hat{\bm n}}$ [as in Fig.~\ref{Fig.2}(c)]. It is known that the number of electrons $q$ pumped through the system per cycle is given by the winding number of Zak phase~\cite{PhysRevB.27.6083}
\begin{eqnarray}\label{Zak}
\gamma(\phi_{\hat{\bm n}})=\int_{-\pi}^{\pi} \mathcal A_k\,dk,
\end{eqnarray}
when $\phi_{\hat{\bm n}}$ increases by $2\pi$. Here, $\mathcal A_k$ is the Berry connection for the valence band at a fixed $\phi_{\hat{\bm n}}$. Equivalently, $q$ also corresponds to the Chern number of valence band in $k$-$\phi_{\hat{\bm n}}$ space~\cite{Vanderbilt_2018}, i.e.,
\begin{eqnarray}\label{Chern}
q=\frac{1}{2\pi}\int \Omega_{k\phi_{\hat{\bm n}}}\,dkd\phi_{\hat{\bm n}},
\end{eqnarray}
where $\Omega_{k\phi_{\hat{\bm n}}}=\partial_k \mathcal A_{\phi_{\hat{\bm n}}}-\partial_{\phi_{\hat{\bm n}}}\mathcal A_k$ is the Berry curvature.

Since this $k$-$\phi_{\hat{\bm n}}$ torus is an \emph{invariant} space of $\prescript{+}{}{\cal T}$, one might naively guess that $q$ must vanish, by the common wisdom that Chern number should vanish for a system preserving time-reversal symmetry. Recall that for a 2D insulator with the usual $\cal T$ symmetry, its Brillouin zone is an invariant space under $\cal T$, with the action ${\cal T}: (k_x, k_y)\mapsto (-k_x,-k_y)$. This action connects the Zak phases $\gamma$ (along $k_x$ direction) at $k_y$ and $-k_y$, such that $\gamma(k_y)=\gamma(-k_y)$. This condition dictates that $\gamma(k_y)$
must have a zero winding number for $k_y$ goes from $-\pi$ to $\pi$, so the Chern number must vanish.

However, the situation here differs at a fundamental point: $\prescript{+}{}{\cal T}$ acts on $k$-$\phi_{\hat{\bm n}}$ space by $\prescript{+}{}{\cal T}: (k, \phi_{\hat{\bm n}})\mapsto (-k, \phi_{\hat{\bm n}}+\pi)$. It behaves like a glide mirror, as illustrated in Fig.~\ref{Fig.2}(d), distinct from the action on conventional 2D Brillouin zone. Hence, a connection is established between Zak phases at $\phi_{\hat{\bm n}}$ and $\phi_{\hat{\bm n}}+\pi$, with
\begin{eqnarray}\label{ZZ}
\gamma(\phi_{\hat{\bm n}})=\gamma(\phi_{\hat{\bm n}}+\pi).
\end{eqnarray}
This condition has two crucial consequences. (i) It no longer forbids a nontrivial winding (Chern) number. (ii) It dictates that the winding pattern of $\gamma$ must be identical for $\phi_{\hat{\bm n}}\in (0,\pi]$ and $(\pi,2\pi]$, so the winding (Chern) number must be an \emph{even} integer, i.e., $q\in 2\mathbb Z$. These points are explicitly confirmed in Fig.~\ref{Fig.2}(e), showing the result of the EMH in Eq.~(\ref{H2D}).

We emphasize that the finding of $2\mathbb Z$ charge pumping is a \emph{general} feature, not limited to the simple model (\ref{H2D}). As argued above, it follows solely from the $\prescript{+}{}{\cal T}$ symmetry (which is always present in ESG) and its \emph{unique action} on the defining space of EMH.

Moreover, other crystal symmetries may impose further constraints on charge pumping, due to their distinct actions on
$k$-$\phi_{\hat{\bm n}}$ space. For example, a $p$-fold rotation acts like a \emph{fractional} translation along $\phi_{\hat{\bm n}}$, since $\prescript{+}{}{C_{pz}}:(k, \phi_{\hat{\bm n}})\mapsto (k, \phi_{\hat{\bm n}}+2\pi/p)$. This means the fundamental domain for EMH is only $1/p$ of the original $k$-$\phi_{\hat{\bm n}}$ torus, {as illustrated in Fig.~\ref{Fig.2}(f)}, which further requires $q\in p\mathbb Z$.
Specifically, in the EMH (\ref{H2D}), replacing $\prescript{+}{}{C_{2z}}$ by $\prescript{+}{}{C_{pz}}$ will simply change the arguments of sine and cosine functions to $p'\pi/2$ , with $p'=\text{lcm}(2,p)$. And the combined effect of $\prescript{+}{}{\cal T}$ and $\prescript{+}{}{C_{pz}}$ will make $q$ valued in $p'\mathbb Z$. For example,
having $\prescript{+}{}{C_{3z}}$ symmetry would quantize $q$ in multiples of six.

\begin{figure}
	\includegraphics[width=8 cm]{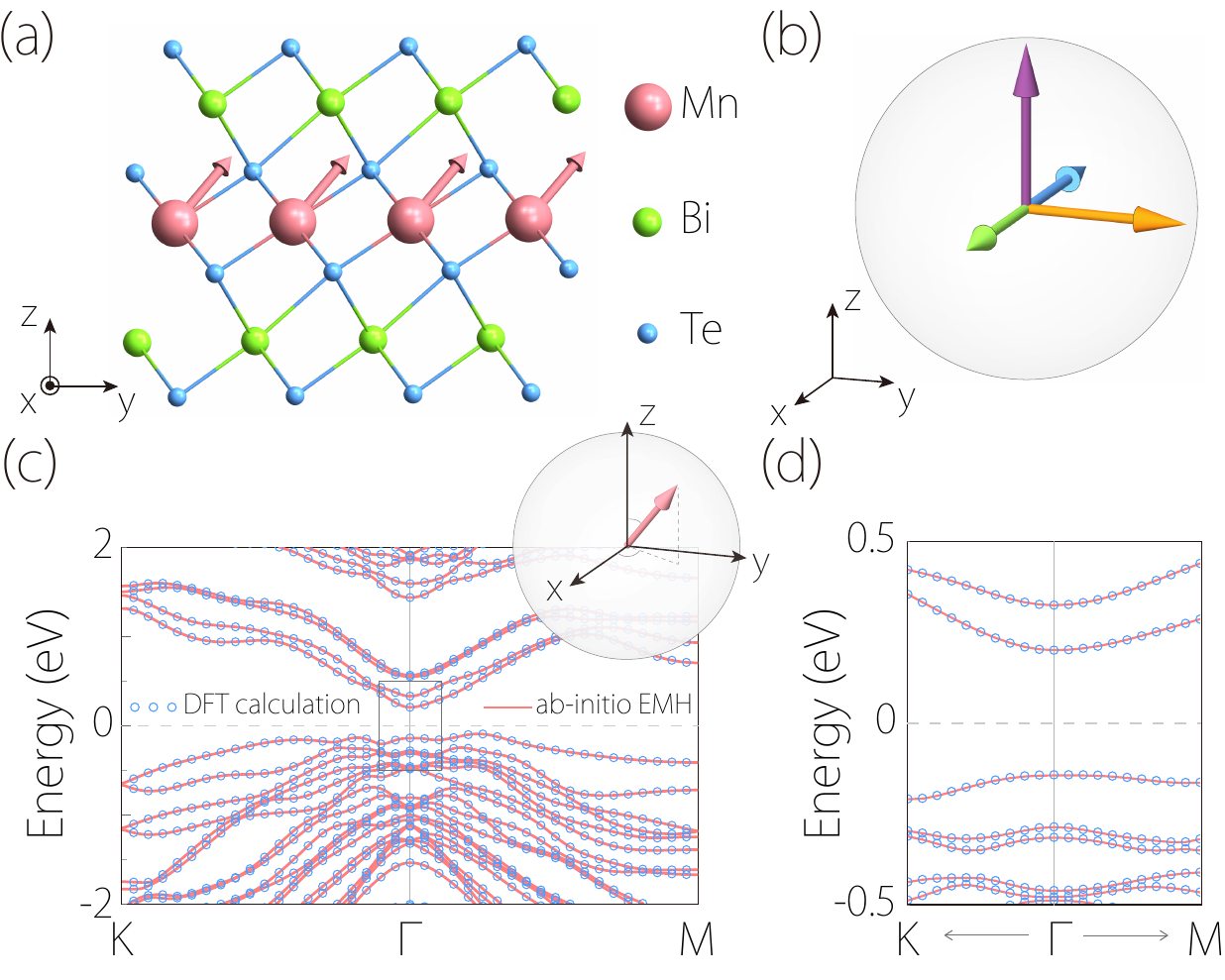}
	\caption{
(a) Structure of monolayer MnBi$_2$Te$_4$. (b) Four different $\hat{\bm n}$ directions ($\pm\hat x$, $\hat y$, and $\hat z$) used in the DFT calculations for extracting the $\mathcal U$ matrices. These $\mathcal U$ matrices are used to construct an \emph{ab-initio} EMH. (c)
Comparison between band structures obtained from this \emph{ab-initio} EMH (red lines) and from direct
DFT calculation (blue circles), for the magnetization direction \((\theta_{\hat{\bm n}},\phi_{\hat{\bm n}})=(\pi/4,\pi/3)\) , illustrated in the inset. (d) Enlarged view of the boxed region in (c).
		}
	\label{Fig.3}
\end{figure}

\textit{\textcolor{blue}{Discussion.}}
We have developed a general approach for constructing EMHs, which establishes the basis for studying physics associated with the direction of magnetic order parameter and with magnetic dynamics. We showed that the unusual topology of EMH's defining space and the distinct symmetry action can lead to remarkable consequences.
This opens the door to a realm of fascinating phenomena, as already demonstrated in a simple example.

Particularly, the added $\hat{\bm n}$ degree of freedom in EMH offers a simple route to interesting topological structures in higher dimensions. As an example, we demonstrate that a 3D antiferromagnetic EMH can realize a topological pump of surface anomalous Hall conductivity, characterized by a second Chern number~\cite{Vanderbilt_2018} {(see End Matter)}.

Besides effective models, our approach can also greatly facilitate first-principles studies of real materials. The idea can be implemented in two ways. The first way is to directly use our approach to construct a tight-binding EMH, using the local orbitals that are of interest, and then obtain parameters in the model by fitting the first-principles band structures. This should work well when the number of bands involved is small (e.g., a few low-energy bands near Fermi level). An alternative way is inspired by the expansion in Eq.~(\ref{HamExpand}). One first constructs Wannier tight-binding models for several $\hat{\bm n}$ directions, using first-principles DFT calculations; then uses these models to directly extract $\cal U$ matrices by taking the lowest few $l$ terms in Eq.~(\ref{HamExpand}). For example, limited to $l=1$, we have Eq.~(\ref{77}), and a calculation of four $\hat{\bm n}$ directions is sufficient for extracting the four $\cal U$ matrices in Eq.~(\ref{77}).
We apply this method to construct an \emph{ab-initio} EMH for monolayer MnBi$_2$Te$_4$ {[Fig.~\ref{Fig.3}(a,b)]}~\cite{SM,C3CE40643A,PredictionAndObservationMnBiTe}. As shown in Fig.~\ref{Fig.3}(c) and (d), the model is surprisingly accurate, achieving excellent agreement with DFT band structure for arbitrary $\hat{\bm n}$ direction.

Finally, our approach can be naturally extended to non-collinear magnetic systems~\cite{Noncollinear2020,Noncollinear2025}. In the most general case, the local moments at magnetic sites are all independent. Label their moment directions with $\hat{\bm m}_1, \cdots, \hat{\bm m}_N$, with $N$ the number of magnetic sublattices. Under operations in $G_0$, these magnetic sublattices are divided into orbits, i.e., two sublattices in the same orbit are related by some symmetry in $G_0$ (when we neglect magnetism).
Since the extension to multiple orbits is straightforward, here, let's assume there is only one orbit, meaning that any two sublattices are connected by some $X\in G_0$.
Then, each element of ESG $\cal G$ has the form $\prescript{\zeta}{}X$, where $\zeta\in S_N$ is an element of the symmetric group, determined by how $X$ permutes the $N$ sublattices. And the constraint relation now takes the form of
\begin{eqnarray}\label{GG}
	D(\prescript{\zeta}{}X)&&{\cal H}({\bm k}, \hat{\bm m}_1, \cdots, \hat{\bm m}_N)D(\prescript{\zeta}{}X)^{-1}\nonumber \\&&= {\cal H}(X{\bm k}, X\hat{\bm m}_{\zeta^{-1}(1)}, \cdots, X\hat{\bm m}_{\zeta^{-1}(N)}).
\end{eqnarray}
Such non-collinear EMHs would be an interesting topic to explore in future studies.

\bibliography{ref}

\clearpage

\appendix*
\setcounter{equation}{0}
\section{End Matter}\label{sec:EM}

\begin{figure}[b]
	\includegraphics[width=8 cm]{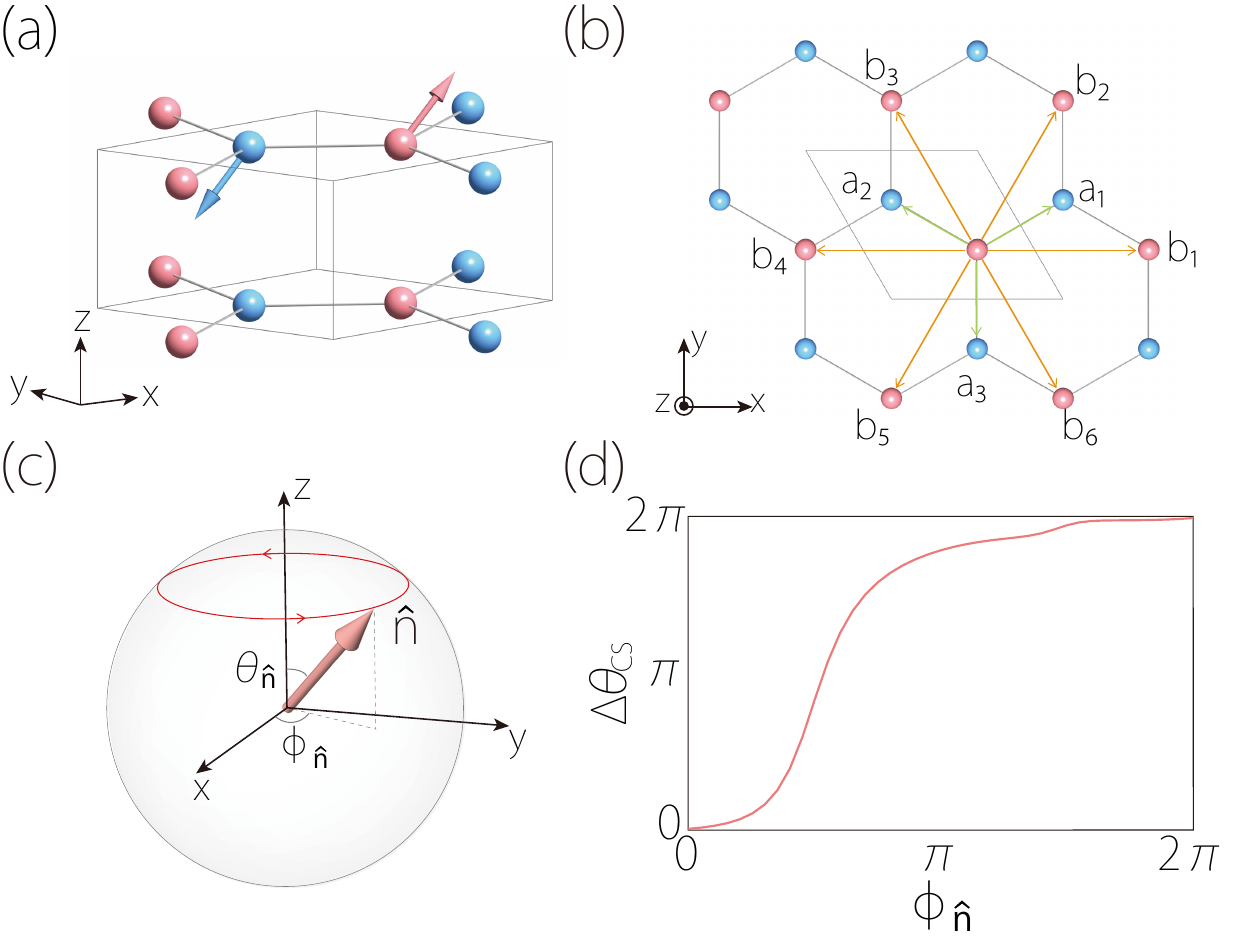}
	\caption{(a) The 3D model consists of 2D honeycomb lattice stacked with AA-stacking order. The two sublattices have opposite magnetic moments, forming an antiferromagnetic order. (b) The intralayer hopping vectors $\boldsymbol{a}_n$ and $\boldsymbol{b}_n$. (c) We consider the pumping induced by the precession of $\hat{\bm n}$ with a fixed $\theta_{\hat{\bm n}}=\pi/4$. (d) Evolution of Chern-Simons angle $\theta_{\mathrm{CS}}$ as $\phi_{\hat{\bm n}}$ varies from $0$ to $2\pi$, exhibiting a nontrivial winding. In the calculation, we take $c_1=1~\mathrm{eV}$, $c_2=-4~\mathrm{eV}$, $c_3=1.63~\mathrm{eV}$, $c_4=0.33~\mathrm{eV}$, $c_5=4.24~\mathrm{eV}$, $c_6=0.82~\mathrm{eV}$, $c_7=0.21~\mathrm{eV}$, and $c_8=1.79~\mathrm{eV}$.
	}
	\label{Fig.4}
\end{figure}

\textit{\textcolor{blue}{3D antiferromagnetic model \& second Chern number.}}
EMH allows access to nontrivial topology associated with higher-dimensional spaces.
Consider a 3D AA-stacked honeycomb antiferromagnetic lattice, with two active sites per unit cell carrying opposite magnetic moments [see Fig.~\ref{Fig.4}(a)]. Each site has two \(s\)-like spin-polarized orbitals \(\ket{\uparrow}\) and \(\ket{\downarrow}\), so there are totally four basis orbitals in a unit cell.
Take ESG generators as $\prescript{-}{}{\cal{P}}$ and ${\prescript{+}{}{\cal{T}}}$, whose matrix representations are
\begin{equation}
	D(\prescript{-}{}{\cal{P}})=\sigma_0\otimes\tau_1,D({\prescript{+}{}{\cal{T}}})=-i\sigma_2\otimes\tau_0 K,
\end{equation}
where $\sigma$ and $\tau$ are Pauli matrices acting on the spin and sublattice spaces, respectively.
Up to $l=2$, the EMH has the form of
\begin{eqnarray} \label{H2C}
	{\cal H}&= {\cal{U}}_{01}Y_{01}  +\sum_{j=1}^3 {\cal U}_{1j}Y_{1j}(\theta_{\hat{\bm n}},\phi_{\hat{\bm n}})\\&+\sum_{j=1}^5 {\cal U}_{2j}Y_{2j}(\theta_{\hat{\bm n}},\phi_{\hat{\bm n}}).  \notag
\end{eqnarray}
The explicit forms of $Y_{lm}$ are shown in SM~\cite{SM}.
Based on Eq.~(\ref{HknConstrain}), we take the following symmetry-allowed $\cal{U}$ matrices for the EMH, up to second-neighbor intralayer hopping and nearest-neighbor interlayer hopping:
\begin{equation}
	\begin{gathered}\label{SecondChernU}
		{\cal U}_{01}
		=
		\left[
		\begin{array}{cc}
			d(\boldsymbol{k})\sigma_3 & f(\boldsymbol{k})\sigma_0\\
			f^*(\boldsymbol{k})\sigma_0 & -d(\boldsymbol{k})\sigma_3
		\end{array}
		\right],
		\\
		{\cal U}_{11}
		=
		c_3\,\sigma_3\otimes\tau_3,
		\qquad
		{\cal U}_{12}
		=
		A_2\,\sigma_1\otimes\tau_3,
		\\
		{\cal U}_{13}
		=
		\left(c_5\sigma_3+A_3\sigma_1\right)\otimes\tau_3,
		\qquad
		{\cal U}_{21}
		=
		{\cal U}_{24}
		=
		{\cal U}_{25}
		=
		0,
		\\
		{\cal U}_{22}
		=
		-c_7\sin k_z\,\sigma_2\otimes\tau_3,
		\qquad
		{\cal U}_{23}
		=
		-c_8\sin k_z\,\sigma_2\otimes\tau_3.
	\end{gathered}
\end{equation}
where $c_i$'s are real model parameters, $	d(\boldsymbol{k})=-ic_1\sum_{n=1}^{6}(-1)^n e^{-i\boldsymbol{b}_n\cdot\boldsymbol{k}}$,
$f(\boldsymbol{k})=c_2\sum_{n=1}^{3}e^{-i\boldsymbol{a}_n\cdot\boldsymbol{k}}
$, $	A_2=
c_4(\cos k_z-1)$, $	A_3
=
c_6(\cos k_z+1)$, $\boldsymbol{a}_n$ and $\boldsymbol{b}_n$ are nearest-neighbor and second-neighbor hopping vectors, as shown in Fig.~\ref{Fig.4}(b). Note that Eq.~(\ref{SecondChernU}) retains only the necessary terms for having nontrivial topology, while coefficients for inessential terms are set to zero for simplicity.

When the magnetic moments precess at a fixed polar angle $\theta_{\hat{\bm n}}$, as shown in Fig.~\ref{Fig.4}(c), the parameter space of the EMH in Eq.~\eqref{H2C} becomes four-dimensional, namely $(k_x,k_y,k_z,\phi_{\hat{\bm n}})$, over which a second Chern number ${\cal C}_{\bm k,\phi_{\hat{\bm n}}}^{(2)}$ is defined:
\begin{equation}
	{\cal C}_{k, \phi_{\hat{\bm n}}}^{(2)}
	=
	\frac{1}{32 \pi^2}
	\int_0^{2\pi} d\phi_{\hat{\bm n}}
	\int_{\text{BZ}} d^3 k\,
	\epsilon_{abcd}\,
	\mathrm{Tr}\!\left[\Omega_{ab}\Omega_{cd}\right],
	\label{C2}
\end{equation}
where $\epsilon_{abcd}$ is the Levi--Civita symbol, $\Omega_{ab}$ denotes the Berry-curvature tensor in the $ab$ subspace, the indices $a,b,c$ and $d$ run over the four coordinates $(k_x,k_y,k_z,\phi_{\hat{\bm n}})$.

Analogous to the fact that the first Chern number characterizes the winding of the Zak phase, the second Chern number characterizes the winding of the Chern-Simons angle $\theta_\text{CS}$ during a pumping cycle~\cite{PhysRevLett.114.096401,PhysRevB.95.075137}. $\theta_\text{CS}$ is defined as
\begin{equation}
	\theta_{\mathrm{CS}}
	=
	-\frac{1}{4\pi}
	\int_{\mathrm{BZ}}
	\varepsilon_{abc}\,
	\operatorname{Tr}\!\left[
	\mathcal A_a \partial_b \mathcal A_c
	- i\frac{2}{3} \mathcal A_a \mathcal A_b \mathcal A_c
	\right]\, d^3k,
	\label{theta_CS}
\end{equation}
where $\partial_a\equiv \partial/\partial k_a$, and $\mathcal A_a$ is the Berry connection along the  direction $a$ in $k$-space.
After one cycle,
\begin{eqnarray}
  \Delta\theta_\text{CS}=2\pi{\cal C}_{k, \phi_{\hat{\bm n}}}^{(2)}.
\end{eqnarray}
Physically, this has been interpreted as a pumping of the surface anomalous Hall conductivity.

For our current EMH, in Fig.~\ref{Fig.4}(d), we plot the variation~\cite{SpinPrecession}
$
	\Delta\theta_{\mathrm{CS}}(\phi_{\hat{\bm n}})
	\equiv
	\theta_{\mathrm{CS}}(\phi_{\hat{\bm n}})
	-
	\theta_{\mathrm{CS}}(0)
$
during the evolution of $\phi_{\hat{\bm n}}$ from $0$ to $2\pi$ ($\theta_{\hat{\bm n}}$ is fixed at $\pi/4$), which exhibits a nontrivial winding pattern.
This corresponds to the case with ${\cal C}_{k,\phi_{\hat{\bm n}}}^{(2)}=1$.

\end{document}


\title{Supplemental Material for ``Equivariant Space Group and Hamiltonian for Collinear Magnetic Systems''}
\author{Chaoxi Cui}
\affiliation{Centre for Quantum Physics, Key Laboratory of Advanced Optoelectronic
	Quantum Architecture and Measurement (MOE), School of Physics, Beijing
	Institute of Technology, Beijing 100081, China}
\affiliation{Beijing Key Lab of Nanophotonics \& Ultrafine Optoelectronic Systems,
	School of Physics, Beijing Institute of Technology, Beijing 100081,
	China}

\author{Zhi-Ming Yu}
\email{zhiming\_yu@bit.edu.cn}
\affiliation{Centre for Quantum Physics, Key Laboratory of Advanced Optoelectronic
	Quantum Architecture and Measurement (MOE), School of Physics, Beijing
	Institute of Technology, Beijing 100081, China}
\affiliation{Beijing Key Lab of Nanophotonics \& Ultrafine Optoelectronic Systems,
	School of Physics, Beijing Institute of Technology, Beijing 100081,
	China}

\author{Yilin Han}
\affiliation{Centre for Quantum Physics, Key Laboratory of Advanced Optoelectronic
	Quantum Architecture and Measurement (MOE), School of Physics, Beijing
	Institute of Technology, Beijing 100081, China}
\affiliation{Beijing Key Lab of Nanophotonics \& Ultrafine Optoelectronic Systems,
	School of Physics, Beijing Institute of Technology, Beijing 100081,
	China}

\author{Run-Wu Zhang}
\affiliation{Centre for Quantum Physics, Key Laboratory of Advanced Optoelectronic
	Quantum Architecture and Measurement (MOE), School of Physics, Beijing
	Institute of Technology, Beijing 100081, China}
\affiliation{Beijing Key Lab of Nanophotonics \& Ultrafine Optoelectronic Systems,
	School of Physics, Beijing Institute of Technology, Beijing 100081,
	China}

\author{Shengyuan A. Yang}
\email{shengyuan.yang@polyu.edu.hk}
\affiliation{Research Laboratory for Quantum Materials, Department of Applied Physics,
	The Hong Kong Polytechnic University, Kowloon, Hong Kong, China}

\author{Yugui Yao}
\email{ygyao@bit.edu.cn}
\affiliation{Centre for Quantum Physics, Key Laboratory of Advanced Optoelectronic
	Quantum Architecture and Measurement (MOE), School of Physics, Beijing
	Institute of Technology, Beijing 100081, China}
\affiliation{Beijing Key Lab of Nanophotonics \& Ultrafine Optoelectronic Systems,
	School of Physics, Beijing Institute of Technology, Beijing 100081,
	China}
\affiliation{Beijing Institute of Technology, Zhuhai 519000, China}

\maketitle

\tableofcontents

\section{Deriving  equivariant space group from  spin space group}
In the main text, the  equivariant space groups (ESGs) and the ESG-operation ${}^{\xi}X$ is obtained from the space group $G_0$ of the corresponding
nonmagnetic lattice. Here, we demonstrate that each ESG actually is isomorphic to a quotient group of corresponding spin space group (SSG).
Consequently, the   ESG $\cal{G}$ can be derived from the SSG $G_S$.

For a collinear magnetic structure with one or two magnetic sublattices, the local magnetic moment on  site $\alpha$ can be written as
\begin{equation}
	\bm m_\alpha=\eta_\alpha m\hat{\bm n},
	\qquad
	\eta_\alpha=\pm 1 ,
\end{equation}
where $\hat{\bm n}$ is the magnetic order parameter orientation and $m=|\bm m|$ is the magnitude of $\bm m$.
$\eta_\alpha$ is always the  same ($\eta_\alpha=1$) for the ferromagnetic systems  with only one  magnetic sublattice, and $\eta_\alpha=\pm1$ label the two spin sublattices in antiferromagnetic systems  with two magnetic sublattices.
Thus, the sites with same $\eta_\alpha$ belong to the same magnetic sublattice, while the sites with different  $\eta_\alpha$ belong to opposite magnetic sublattices.

The classification of the   magnetic sublattice  can be  more directly observed in  the framework of SSG. An element  of SSG is written as $[U||X]$~\cite{Xiao2024PRXSSG,Chen2024PRXSSG,Jiang2024PRXSSG,LiuQihangNature},
where $X$ is a real-space operation and $U$ is the accompanying spin-space operation. If $X$ maps site $\alpha$ to site $\beta$, the invariance of the
collinear magnetic structure requires
\begin{equation}
	\eta_\beta \hat{\bm n}
	=
	U\eta_\alpha \hat{\bm n}= \eta_\alpha U \hat{\bm n}.
\end{equation}
Therefore, if
\begin{equation}
	U\hat{\bm n}=+\hat{\bm n},
\end{equation}
then one has $\eta_\beta=\eta_\alpha$, indicating that  $X$ preserves the magnetic  sublattice.
In contrast, if
\begin{equation}
	U\hat{\bm n}=-\hat{\bm n},
\end{equation}
then $\eta_\beta=-\eta_\alpha$, which indicates that $X$ switches the two spin sublattices.
Thus, the preserving or switching character of a real-space operation is
determined by the spin part of the corresponding SSG operation.

This SSG construction is consistent with the construction of the ESG used in the main text.
Explicitly, the spatial part \(X\) of any SSG operation \([U||X]\) must be  a symmetry operation of the corresponding nonmagnetic lattice, and hence belongs to \(G_0\).
When $X$ preserves (switches) the magnetic  sublattices, it appears in the SSG as  \([E||X]\) $([C_{2}||X])$, where $E$ is the spin identity operation and
$C_{2}$ flips the spin.
Similarly, for any operation in ESG: $^\xi X$, one has $\xi=1$ ($\xi=-1$) for $X$ preserves (switches) the magnetic  sublattices.

Next, we demonstrate  the group-theoretical relation between the SSG and our ESG. For a collinear magnet, the SSG can be decomposed as~\cite{Xiao2024PRXSSG,Chen2024PRXSSG,Jiang2024PRXSSG,LiuQihangNature}
\begin{equation}
	{G_S}=({G}_{NS}\times Z_2^K)\ltimes{{\cal S}_0},
\end{equation}
where ${G}_{NS}$ denotes the nontrivial (spin-only-free) part of the SSG, and
\begin{equation}
	Z_2^K=\{[E||E],[C_{2}{\cal T}||E]\}.
\end{equation}
The operation $C_{2}{\cal T}$ leaves the collinear order parameter unchanged, because ${\cal T}$ reverses $\hat{\bm n}$ and $C_{2}$ reverses it back.
Besides
\begin{equation}
	{{\cal S}_0}=\{[C_{\infty}||E]\},
\end{equation}
is the trivial spin-only subgroup, where $C_{\infty}$ denotes an
arbitrary spin rotation along $\hat{\bm n}$.
Since \({\cal S}_0\) does not change the collinear magnetic structure, the ESG can be obtained by taking the  quotient of the SSG by \({\cal S}_0\).
Explicitly, the corresponding quotient group is written as
\begin{equation}
	{G_S}/{{\cal S}_0} 	\simeq 	{G}_{NS}\times Z_2^K.
\end{equation}
Remarkably, for this quotient group, the spin part of a unitary operation has only two choices: $E$, which leaves $\hat{\bm n}$ invariant, and
$C_{2}$, which flips $\hat{\bm n}$.
As a consequent,  the correspondence between the ${G_S}/{{\cal S}_0} $ and the ESG  can be written as
\begin{equation} \label{S9}
	[E||X]\rightarrow{}^{+}X,
	\qquad
	[C_{2}||X]\rightarrow{}^{-}X .
\end{equation}
The first operation preserves the spin sublattices, whereas the second
switches the two spin sublattices. Besides, the antiunitary operation  gives
\begin{equation}
	[C_{2}{\cal T}||E]\rightarrow{}^{+}{\cal T}.
\end{equation}
Here, the superscript $+$ in ${}^{+}{\cal T}$ indicates that pure time reversal does not exchange the spin sublattices in real space. More generally, if an antiunitary operation contains an additional real-space operation $X$, its superscript is still determined by whether this real-space operation preserves or switches the spin sublattices.
All these result demonstrate that the ESG  ($\cal G$) is isomorphic to a quotient group of SSG,
\begin{equation}\label{S11}
	{\cal G} \simeq{G_S}/{{\cal S}_0} 	\simeq 	{G}_{NS}\times Z_2^K.
\end{equation}

Due to the closed relationship  between $\cal G$ and ${G_S}/{{\cal S}_0}$, each ESG and its operations can be obtained from the corresponding SSG.
In practice, the ESG of a collinear magnet can be obtained from its SSG in three steps:
\begin{enumerate}[label=(\arabic*), leftmargin=3em]
	\item Identifying the SSG ${G_S}$ of the collinear magnet.
	\item Taking the   quotient of the SSG ${G_S}$ by \({\cal S}_0\).
	\item Replacing  each unitary operation $[U||X]$ in ${G_S}/{{\cal S}_0}$ by ${}^{+}X$ (${}^{-}X$) if $U=E$ ($U=C_2$), and replacing $[C_2{\cal{T}}||E]$ by $^+{\cal{T}}$. The other antiunitary operations can be generated by the  product of $^+{\cal{T}}$ and ${}^{\xi}X$ (with $\xi=\pm$).
\end{enumerate}

To make the above correspondence explicit, we consider a simple example shown in Fig. 1 in the main text.
Suppose the SSG of a collinear magnetic system is
\begin{equation}
	{G_S}
	=
	\left(
	\{
	[E||E],\,
	[C_{2}||C_4],\,
	[E||C_2],\,
	[C_{2}||C_4^3]
	\}
	\times Z_2^K
	\right)
	\ltimes {{\cal S}_0}.
\end{equation}
The translational part of the SSG is omitted here, since it acts trivially on the magnetic configuration.  Taking the quotient of the SSG by the trivial spin-only subgroup
\({{\cal S}_0}\), one obtains
\begin{equation}
	{G_S}/{{\cal S}_0}
	\simeq
	\{
	[E||E],\,
	[C_{2}||C_4],\,
	[E||C_2],\,
	[C_{2}||C_4^3]
	\}
	\times
	\{
	[E||E],\,
	[C_{2}{\cal T}||E]
	\}.
\end{equation}
Using the aforementioned correspondence  [see Eqs.~(\ref{S9}-\ref{S11})], the unitary part of the corresponding  ESG can be obtained  as
\begin{equation}
	{\cal G}_{\rm u}
	=
	\{
	{}^{+}E,\,
	{}^{-}C_4,\,
	{}^{+}C_2,\,
	{}^{-}C_4^3
	\}.
\end{equation}
Including the time-reversal sector gives the full ESG, expressed as
\begin{equation}
	{\cal G}
	=
	{\cal G}_{\rm u}\times\{{}^{+}E,{}^{+}{\cal T}\}=\{
	{}^{+}E,\,
	{}^{-}C_4,\,
	{}^{+}C_2,\,
	{}^{-}C_4^3,\,
	{}^{+}{\cal T},\,
	{}^{-}C_4{\cal T},\,
	{}^{+}C_2{\cal T},\,
	{}^{-}C_4^3{\cal T}
	\}.
\end{equation}
This example shows explicitly how the spin part of a SSG operation determines the superscript $\xi$ of the corresponding ESG operation.

\section{Real spherical harmonics and  $S$ matrices}
In this section, we give the explicit forms of the real spherical harmonics \(Y_{lm}(\theta_{\hat{\bm n}},\phi_{\hat{\bm n}})\) used in the main text, and the  $S_l(X)$ matrices for typical operators.

The real spherical harmonics can be obtained from the conventional complex spherical harmonics $Y_l^m$.
Here, we set
\begin{equation}
	\begin{aligned}\label{SH0}
Y_{01} &=\frac{1}{2} \sqrt{\frac{1}{\pi}},
	\end{aligned}
\end{equation}
for $l=0$,
\begin{equation}
	\begin{aligned} \label{SH1}
	Y_{11} &= \frac{1}{2}\sqrt{\frac{3}{\pi}} \sin\!\left(\theta_{\hat{\bm n}}\right)\cos\!\left(\phi_{\hat{\bm n}}\right),\\
	Y_{12} &= \frac{1}{2}\sqrt{\frac{3}{\pi}} \sin\!\left(\theta_{\hat{\bm n}}\right)\sin\!\left(\phi_{\hat{\bm n}}\right),\\
	Y_{13} &= \frac{1}{2}\sqrt{\frac{3}{\pi}} \cos\!\left(\theta_{\hat{\bm n}}\right),
	\end{aligned}
\end{equation}
for $l=1$,
and
\begin{equation}
	Y_{lm}
	=
	\begin{cases}
		\dfrac{(-1)^{\mu}}{\sqrt{2}}
		\left[
		Y_l^{\mu}
		+
		\left(Y_l^{\mu}\right)^*
		\right],
		& \mu>0, \\[12pt]
		Y_l^0,
		& \mu=0, \\[12pt]
		\dfrac{(-1)^{\mu}}{i\sqrt{2}}
		\left[
		Y_l^{|\mu|}
		-
		\left(Y_l^{|\mu|}\right)^*
		\right],
		& \mu<0 .
	\end{cases}
\end{equation}
for $l>1$, where $\mu=m-l-1$ and $m=1,2,\cdots2l+1$.

Then, we discuss the expression of $S_l$ matrix. 
Notice that for an operator $^\xi X$ in ESG, the expression of $S_l$ matrix is determined by the operator $X$  and is irrelevant to $\xi$. Thus, we define 
\begin{equation}	
S_l(X)\equiv S_l(^\xi X).
\end{equation}
For $l=0$, since \(Y_{01}\) is a constant, one has \(S_{0}=1\) for all operators.
The \(S\) matrices for \(l=1,2,\) and \(3\) are listed in Table~\ref{SMatrix}.
In Table~\ref{SMatrix}, we present only the \(S\) matrices for proper operations, because any improper operation can be decomposed into a proper operation \(R\) followed by spatial inversion symmetry  \(\mathcal{P}\).
Moreover, because \(\mathcal{P}\) keeps the magnetic order parameter orientation  $\hat{\bm n}$  invariant, the matrix $S_l(\mathcal{P})$ is always an identity matrix, indicating that \(S(\mathcal{P}R)=S(R)\).

\clearpage

\begin{table*}[t]
	\caption{$S$ matrices for typical proper point group symmetries with $l=1,2$ and $3$.
	The \(S\) matrix for an improper operation \(\mathcal{P}R\) can be obtained from \(S(\mathcal{P}R)=S(R)\), where \(\mathcal{P}\) is spatial inversion.
	 }\label{SMatrix}
	\begin{ruledtabular}
		\setlength{\arraycolsep}{3pt}
		\renewcommand{\arraystretch}{1.05}
		\newcommand{\rowpad}{\rule[-2.8ex]{0pt}{7.4ex}}
\newsavebox{\lthreebox}
\newcommand{\lthree}[1]{%
	\sbox{\lthreebox}{%
		\scalebox{0.85}{$\left[
			\begin{array}{ccccccc}
				#1
			\end{array}
			\right]$}%
	}%
	\raisebox{0pt}%
	[\dimexpr \ht\lthreebox+\ht\lthreebox/15\relax]%
	[\dimexpr \dp\lthreebox+\dp\lthreebox/15\relax]%
	{\usebox{\lthreebox}}%
}
		\begin{tabular}{c|c c c}
			& $l=1$ & $l=2$ & $l=3$\tabularnewline
			\hline
			$C_{2x}\rowpad$ &
			$\left[\begin{array}{ccc}
				1 & 0 & 0\\
				0 & -1 & 0\\
				0 & 0 & -1
			\end{array}\right]$ &
			$\left[\begin{array}{ccccc}
				-1 & 0 & 0 & 0 & 0\\
				0 & 1 & 0 & 0 & 0\\
				0 & 0 & 1 & 0 & 0\\
				0 & 0 & 0 & -1 & 0\\
				0 & 0 & 0 & 0 & 1
			\end{array}\right]$ &
			$\lthree{
				-1 & 0 & 0 & 0 & 0 & 0 & 0\\
				0 & 1 & 0 & 0 & 0 & 0 & 0\\
				0 & 0 & -1 & 0 & 0 & 0 & 0\\
				0 & 0 & 0 & -1 & 0 & 0 & 0\\
				0 & 0 & 0 & 0 & 1 & 0 & 0\\
				0 & 0 & 0 & 0 & 0 & -1 & 0\\
				0 & 0 & 0 & 0 & 0 & 0 & 1
			}$\tabularnewline
			\hline
			$C_{2y}\rowpad$ &
			$\left[\begin{array}{ccc}
				-1 & 0 & 0\\
				0 & 1 & 0\\
				0 & 0 & -1
			\end{array}\right]$ &
			$\left[\begin{array}{ccccc}
				-1 & 0 & 0 & 0 & 0\\
				0 & -1 & 0 & 0 & 0\\
				0 & 0 & 1 & 0 & 0\\
				0 & 0 & 0 & 1 & 0\\
				0 & 0 & 0 & 0 & 1
			\end{array}\right]$ &
			$\lthree{
				1 & 0 & 0 & 0 & 0 & 0 & 0\\
				0 & 1 & 0 & 0 & 0 & 0 & 0\\
				0 & 0 & 1 & 0 & 0 & 0 & 0\\
				0 & 0 & 0 & -1 & 0 & 0 & 0\\
				0 & 0 & 0 & 0 & -1 & 0 & 0\\
				0 & 0 & 0 & 0 & 0 & -1 & 0\\
				0 & 0 & 0 & 0 & 0 & 0 & -1
			}$\tabularnewline
			\hline
			$C_{2z}\rowpad$ &
			$\left[\begin{array}{ccc}
				-1 & 0 & 0\\
				0 & -1 & 0\\
				0 & 0 & 1
			\end{array}\right]$ &
			$\left[\begin{array}{ccccc}
				1 & 0 & 0 & 0 & 0\\
				0 & -1 & 0 & 0 & 0\\
				0 & 0 & 1 & 0 & 0\\
				0 & 0 & 0 & -1 & 0\\
				0 & 0 & 0 & 0 & 1
			\end{array}\right]$ &
			$\lthree{
				-1 & 0 & 0 & 0 & 0 & 0 & 0\\
				0 & 1 & 0 & 0 & 0 & 0 & 0\\
				0 & 0 & -1 & 0 & 0 & 0 & 0\\
				0 & 0 & 0 & 1 & 0 & 0 & 0\\
				0 & 0 & 0 & 0 & -1 & 0 & 0\\
				0 & 0 & 0 & 0 & 0 & 1 & 0\\
				0 & 0 & 0 & 0 & 0 & 0 & -1
			}$\tabularnewline
			\hline
			$C_{3,111}\rowpad$ &
			$\left[\begin{array}{ccc}
				0 & 0 & 1\\
				1 & 0 & 0\\
				0 & 1 & 0
			\end{array}\right]$ &
			$\left[\begin{array}{ccccc}
				0 & 0 & 0 & 1 & 0\\
				1 & 0 & 0 & 0 & 0\\
				0 & 0 & -\frac{1}{2} & 0 & -\sqrt{\frac{3}{4}}\\
				0 & 1 & 0 & 0 & 0\\
				0 & 0 & \sqrt{\frac{3}{4}} & 0 & -\frac{1}{2}
			\end{array}\right]$ &
			$\lthree{
				0 & 0 & 0 & 0 & \sqrt{\frac{15}{16}} & 0 & -\frac{1}{4}\\
				0 & 1 & 0 & 0 & 0 & 0 & 0\\
				0 & 0 & 0 & 0 & -\frac{1}{4} & 0 & -\sqrt{\frac{15}{16}}\\
				-\sqrt{\frac{5}{8}} & 0 & -\sqrt{\frac{3}{8}} & 0 & 0 & 0 & 0\\
				0 & 0 & 0 & -\sqrt{\frac{3}{8}} & 0 & -\sqrt{\frac{5}{8}} & 0\\
				-\sqrt{\frac{3}{8}} & 0 & \sqrt{\frac{5}{8}} & 0 & 0 & 0 & 0\\
				0 & 0 & 0 & \sqrt{\frac{5}{8}} & 0 & -\sqrt{\frac{3}{8}} & 0
			}$\tabularnewline
			\hline
			$C_{4x}\rowpad$ &
			$\left[\begin{array}{ccc}
				1 & 0 & 0\\
				0 & 0 & 1\\
				0 & -1 & 0
			\end{array}\right]$ &
			$\left[\begin{array}{ccccc}
				0 & 0 & 0 & 1 & 0\\
				0 & -1 & 0 & 0 & 0\\
				0 & 0 & -\frac{1}{2} & 0 & -\sqrt{\frac{3}{4}}\\
				-1 & 0 & 0 & 0 & 0\\
				0 & 0 & -\sqrt{\frac{3}{4}} & 0 & \frac{1}{2}
			\end{array}\right]$ &
			$\lthree{
				0 & 0 & 0 & -\sqrt{\frac{5}{8}} & 0 & \sqrt{\frac{3}{8}} & 0\\
				0 & -1 & 0 & 0 & 0 & 0 & 0\\
				0 & 0 & 0 & -\sqrt{\frac{3}{8}} & 0 & -\sqrt{\frac{5}{8}} & 0\\
				\sqrt{\frac{5}{8}} & 0 & \sqrt{\frac{3}{8}} & 0 & 0 & 0 & 0\\
				0 & 0 & 0 & 0 & -\frac{1}{4} & 0 & -\sqrt{\frac{15}{16}}\\
				-\sqrt{\frac{3}{8}} & 0 & \sqrt{\frac{5}{8}} & 0 & 0 & 0 & 0\\
				0 & 0 & 0 & 0 & -\sqrt{\frac{15}{16}} & 0 & \frac{1}{4}
			}$\tabularnewline
			\hline
			$C_{4y}\rowpad$ &
			$\left[\begin{array}{ccc}
				0 & 0 & -1\\
				0 & 1 & 0\\
				1 & 0 & 0
			\end{array}\right]$ &
			$\left[\begin{array}{ccccc}
				0 & -1 & 0 & 0 & 0\\
				1 & 0 & 0 & 0 & 0\\
				0 & 0 & -\frac{1}{2} & 0 & \sqrt{\frac{3}{4}}\\
				0 & 0 & 0 & -1 & 0\\
				0 & 0 & \sqrt{\frac{3}{4}} & 0 & \frac{1}{2}
			\end{array}\right]$ &
			$\lthree{
				\frac{1}{4} & 0 & \sqrt{\frac{15}{16}} & 0 & 0 & 0 & 0\\
				0 & -1 & 0 & 0 & 0 & 0 & 0\\
				\sqrt{\frac{15}{16}} & 0 & -\frac{1}{4} & 0 & 0 & 0 & 0\\
				0 & 0 & 0 & 0 & -\sqrt{\frac{3}{8}} & 0 & \sqrt{\frac{5}{8}}\\
				0 & 0 & 0 & \sqrt{\frac{3}{8}} & 0 & -\sqrt{\frac{5}{8}} & 0\\
				0 & 0 & 0 & 0 & \sqrt{\frac{5}{8}} & 0 & \sqrt{\frac{3}{8}}\\
				0 & 0 & 0 & -\sqrt{\frac{5}{8}} & 0 & -\sqrt{\frac{3}{8}} & 0
			}$\tabularnewline
			\hline
			$C_{4z}\rowpad$ &
			$\left[\begin{array}{ccc}
				0 & 1 & 0\\
				-1 & 0 & 0\\
				0 & 0 & 1
			\end{array}\right]$ &
			$\left[\begin{array}{ccccc}
				-1 & 0 & 0 & 0 & 0\\
				0 & 0 & 0 & -1 & 0\\
				0 & 0 & 1 & 0 & 0\\
				0 & 1 & 0 & 0 & 0\\
				0 & 0 & 0 & 0 & -1
			\end{array}\right]$ &
			$\lthree{
				0 & 0 & 0 & 0 & 0 & 0 & 1\\
				0 & -1 & 0 & 0 & 0 & 0 & 0\\
				0 & 0 & 0 & 0 & -1 & 0 & 0\\
				0 & 0 & 0 & 1 & 0 & 0 & 0\\
				0 & 0 & 1 & 0 & 0 & 0 & 0\\
				0 & 0 & 0 & 0 & 0 & -1 & 0\\
				-1 & 0 & 0 & 0 & 0 & 0 & 0
			}$\tabularnewline
			\hline
			$C_{2,110}$ &
			$\left[\begin{array}{ccc}
				0 & 1 & 0\\
				1 & 0 & 0\\
				0 & 0 & -1
			\end{array}\right]$ &
			$\left[\begin{array}{ccccc}
				1 & 0 & 0 & 0 & 0\\
				0 & 0 & 0 & -1 & 0\\
				0 & 0 & 1 & 0 & 0\\
				0 & -1 & 0 & 0 & 0\\
				0 & 0 & 0 & 0 & -1
			\end{array}\right]$ &
			$\lthree{
				0 & 0 & 0 & 0 & 0 & 0 & -1\\
				0 & -1 & 0 & 0 & 0 & 0 & 0\\
				0 & 0 & 0 & 0 & 1 & 0 & 0\\
				0 & 0 & 0 & -1 & 0 & 0 & 0\\
				0 & 0 & 1 & 0 & 0 & 0 & 0\\
				0 & 0 & 0 & 0 & 0 & 1 & 0\\
				-1 & 0 & 0 & 0 & 0 & 0 & 0
			}$
			\label{TabI}
		\end{tabular}
	\end{ruledtabular}
\end{table*}
\clearpage

\section{Computation details of 1D ferromagnetic chain}
In  Sec. ``1D ferromagnetic chain \& $2\mathbb Z$ charge pump'' in the main text, we construct an equivariant magnetic lattice Hamiltonian for a 1D ferromagnetic chain. 
In this lattice  model, the generators of $G_0$ can be chosen as $C_{2x}$, $C_{2z}$ and $\cal {T}$. Accordingly the generators of the  ESG are $\prescript{+}{}C_{2x}$, $\prescript{+}{}C_{2z}$ and $\prescript{+}{}{\cal{T}}$. When considering the $s$-like spin-polarized orbits $\ket{\uparrow} $ and $\ket{\downarrow}$, the matrix representations of the generators are
\begin{eqnarray}
D(\prescript{+}{}C_{2x})=-i\sigma_1,  \ D(\prescript{+}{}C_{2z})=i \sigma_3, \ 
D({\prescript{+}{}{\cal{T}}})=-i\sigma_2 { K},
\end{eqnarray}
with $K$  the complex conjugation operator and $\sigma$'s the Pauli matrices. Up to $l=1$, the EMH is given by
\begin{eqnarray}
	{\cal H} = {\cal{U}}_{01}Y_{01} +\sum_{j=1}^3 {\cal U}_{1j}Y_{1j}(\theta_{\hat{\bm n}},\phi_{\hat{\bm n}}).
\end{eqnarray}
The explicit forms of $Y_{lm}$ are listed in  Eq. (\ref{SH0}) and Eq. (\ref{SH1}).
According to Table \ref{SMatrix}, one has   $S_0({C_{2x}})=S_0({C_{2z}})=S_0({\mathcal{T}})=1$ and
\begin{eqnarray}
	S_1(C_{2x})&=&\left[\begin{array}{ccc}
		+1 & 0 & 0\\
		0 & -1 & 0\\
		0 & 0 & -1
	\end{array}\right], \ 
	S_1(C_{2z})=\left[\begin{array}{ccc}
		-1 & 0 & 0\\
		0 & -1 & 0\\
		0 & 0 & +1
	\end{array}\right],\ \nonumber
	S_1({\mathcal{T}})=\left[\begin{array}{ccc}
		-1 & 0 & 0\\
		0 & -1 & 0\\
		0 & 0 & -1
	\end{array}\right]
\end{eqnarray}

By substituting the $S$ matrices into Eq.~(5) in the main text, we can obtain  the symmetry constrain on ${\cal{U}}$.
We take $\prescript{+}{}C_{2x}$ as an example to explicitly demonstrate the symmetry-constraining procedure. By substituting $D(\prescript{+}{}C_{2x})$, $S_0(C_{2x})$ and $S_1(C_{2x})$ into Eq. (5) in the main text,  we obtain
\begin{eqnarray}
	\sigma_1{\cal{U}}_{01}(k)\sigma_1={\cal{U}}_{01}(-k),
\end{eqnarray}
for $l=0$, and 
\begin{eqnarray}
	\sigma_1{\cal{U}}_{11}(k)\sigma_1&=&\eta\sum_{j=1}^{3} S_1(\prescript{+}{}C_{2x})_{j1} {\cal{U}}_{1j}(-k)={\cal{U}}_{11}(-k),\notag \\
	\sigma_1{\cal{U}}_{12}(k)\sigma_1&=&-{\cal{U}}_{12}(-k), \ \sigma_1{\cal{U}}_{13}(k)\sigma_1=-{\cal{U}}_{13}(-k), \notag \\
\end{eqnarray}
for $l=1$.
These equations express the symmetry constrain imposed by $\prescript{+}{}C_{2x}$ to ${\cal{U}}_0$ and ${\cal{U}}_1$.
Similar to the $\prescript{+}{}C_{2x}$ case, we traverse all symmetry generators and obtain the symmetry-allowed $\cal{U}$ matrices  up to nearest-neighboring hoping as
\begin{eqnarray} \label{Ulm}
	{\cal{U}}_{01} &= & t_{0} \cos k+ s_{0} \sin k\, \sigma_{3}, \\ \notag
	{\cal{U}}_{1j} &=&  (M_j + t_{j} \cos k   ) \sigma_{j}+\delta_{j,3} s_{1} \sin k,
\end{eqnarray}
which are exactly the Eqs. (8-9) in the main text.

\section{Construction of  \emph{ab-initio} EMH}
According to Eq.~(4) in the main text, the expansion of EMH up to $l=1$ contains four unknown coefficient matrices, namely ${\cal U}_{01}$, ${\cal U}_{11}$, ${\cal U}_{12}$, and ${\cal U}_{13}$. They can be determined from first-principles calculations with SOC for four different magnetization directions $(\theta_{\hat{\bm n}_i},\phi_{\hat{\bm n}_i})$ with $i=1,2,3,$ and $4$. For each chosen direction, we perform a   first-principles calculation and construct the  Wannier Hamiltonian ${\cal H}_i^{W}$. Then ${\cal U}_{01}$, ${\cal U}_{11}$, ${\cal U}_{12}$, ${\cal U}_{13}$ and ${\cal H}_i^{W}$ satisfy
\begin{equation}
	\boldsymbol{{\cal H}}^{W}=A\boldsymbol{{\cal U}},
\end{equation}
where $\boldsymbol{{\cal H}}^{W}=({\cal H}_1^{W},{\cal H}_2^{W},{\cal H}_3^{W},{\cal H}_4^{W})^{T}$, $\boldsymbol{{\cal U}}=({\cal U}_{01},{\cal U}_{11},{\cal U}_{12},{\cal U}_{13})^{T}$, and
\begin{equation}
	A=\left(\begin{array}{cccc}
		1 & \sin \theta_{\hat{\bm n}_1} \cos \phi_{\hat{\bm n}_1} & \sin \theta_{\hat{\bm n}_1} \sin \phi_{\hat{\bm n}_1} & \cos \theta_{\hat{\bm n}_1} \\
		1 & \sin \theta_{\hat{\bm n}_2} \cos \phi_{\hat{\bm n}_2} & \sin \theta_{\hat{\bm n}_2} \sin \phi_{\hat{\bm n}_2} & \cos \theta_{\hat{\bm n}_2} \\
		1 & \sin \theta_{\hat{\bm n}_3} \cos \phi_{\hat{\bm n}_3} & \sin \theta_{\hat{\bm n}_3} \sin \phi_{\hat{\bm n}_3} & \cos \theta_{\hat{\bm n}_3} \\
		1 & \sin \theta_{\hat{\bm n}_4} \cos \phi_{\hat{\bm n}_4} & \sin \theta_{\hat{\bm n}_4} \sin \phi_{\hat{\bm n}_4} & \cos \theta_{\hat{\bm n}_4}
	\end{array}\right),
\end{equation}
where we drop the normalization factors of spherical harmonics for simplicity.
If the four magnetization directions $\hat{\bm n}_i$ are noncoplanar, then $\det A\neq0$, and thus ${\cal U}_{01}$, ${\cal U}_{11}$, ${\cal U}_{12}$, and ${\cal U}_{13}$ can be obtained as
\begin{equation}
	\boldsymbol{{\cal U}}=A^{-1}\boldsymbol{{\cal H}}^{W}.
\end{equation}

\begin{figure}
	\includegraphics[width=16cm]{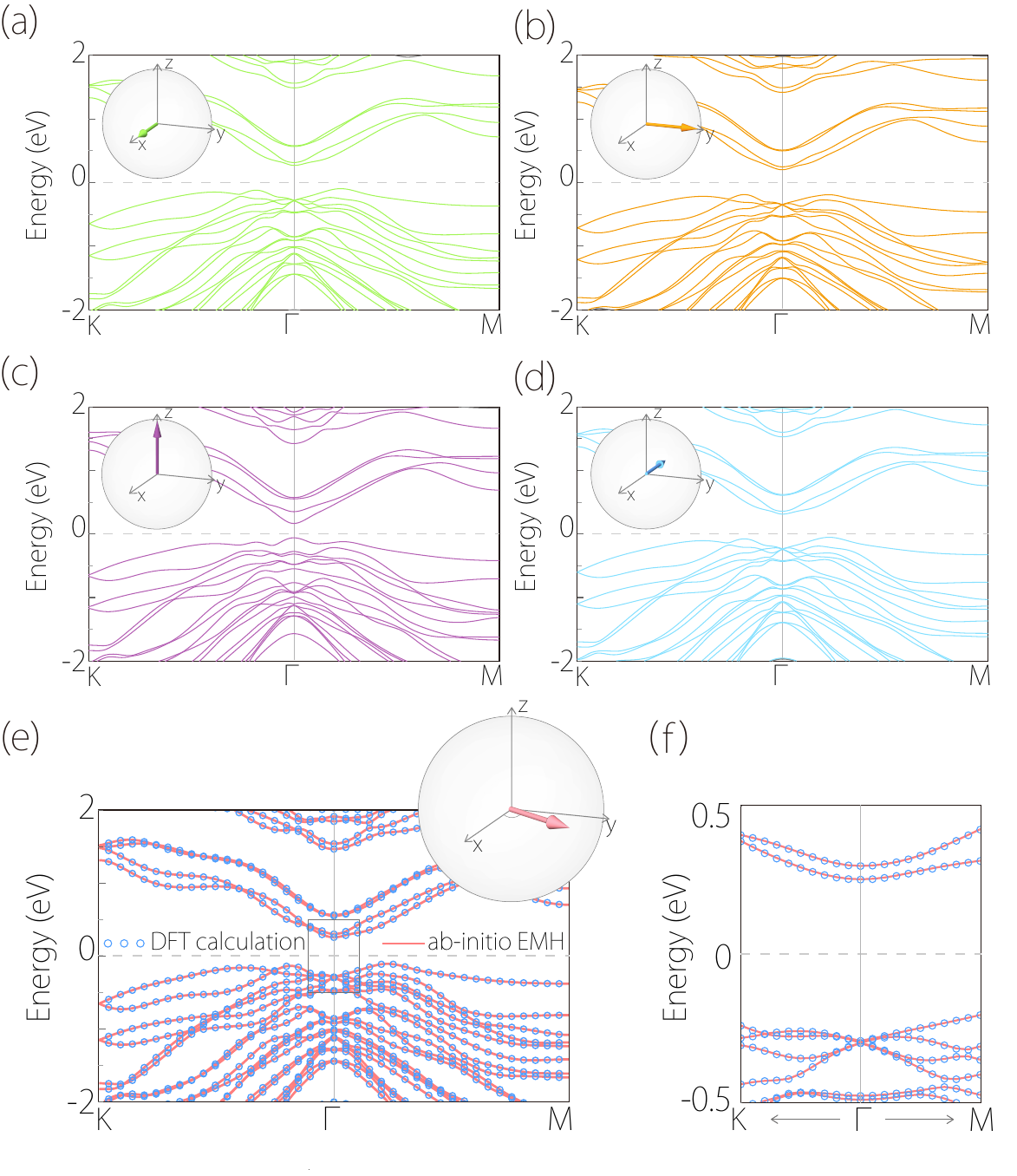}
	\caption{
		(a-d) DFT  band structures of monolayer MnBi$_2$Te$_4$ with SOC for four representative magnetization directions: $(\theta_{\hat{\bm n}_i},\phi_{\hat{\bm n}_i})=(\pi/2,0)$, $(\pi/2,\pi/2)$, $(0,0)$, and $(\pi/2,\pi)$.
		The  direction of \(\hat{\bm n}\)  is shown in the inset of each panel. (e) Comparison between band structures obtained from this \emph{ab-initio} EMH (red lines) and from direct
		DFT calculation (blue circles), for the magnetization direction  \((\theta_{\hat{\bm n}},\phi_{\hat{\bm n}})=(\pi/2,\pi/3)\), illustrated in the inset. (f) Enlarged view of the boxed region in (e).	}
	\label{Fig.5}
\end{figure}

To validate this approach, we perform first-principles calculations for monolayer MnBi$_2$Te$_4$ with four magnetization directions, namely $(\theta_{\hat{\bm n}_i},\phi_{\hat{\bm n}_i})=(\pi/2,0)$, $(\pi/2,\pi/2)$, $(0,0)$, and $(\pi/2,\pi)$.
The corresponding band structures are shown in Fig.~\ref{Fig.5}(a-d), from which one observes that the band structures of monolayer MnBi$_2$Te$_4$ for different magnetization directions differ substantially, due  to the strong SOC. For these four directions, one has 
\begin{equation}
	A=\left(\begin{array}{cccc}
		1 & 1 & 0 & 0 \\
		1 & 0 & 1 & 0 \\
		1 & 0 & 0 & 1 \\
		1 & -1 & 0 & 0
	\end{array}\right),
	\quad
	A^{-1}
	=
	\frac{1}{2}
	\begin{pmatrix}
		1 & 0 & 0 & 1\\
		1 & 0 & 0 & -1\\
		-1 & 2 & 0 & -1\\
		-1 & 0 & 2 & -1
	\end{pmatrix}.
\end{equation}
Accordingly, we obtain
\begin{eqnarray}
	{\cal U}_{01}=\frac{{\cal H}_1^{W}+{\cal H}_4^{W}}{2}&,&\quad
	{\cal U}_{11}={\cal H}_1^{W}-{\cal U}_{01}, \\
	{\cal U}_{12}={\cal H}_2^{W}-{\cal U}_{01}&,&\quad
	{\cal U}_{13}={\cal H}_3^{W}-{\cal U}_{01}. \notag
\end{eqnarray}

To check the validity of the \emph{ab-initio} EMH, we compare the band structure from \emph{ab-initio} EMH  with that obtained from   direct first-principles DFT calculations for two generic  magnetization directions: \((\theta_{\hat{\bm n}},\phi_{\hat{\bm n}})=(\pi/4,\pi/3)\) in Fig.~3(c,d) of the main text and \((\theta_{\hat{\bm n}},\phi_{\hat{\bm n}})=(\pi/2,\pi/3)\) in Fig.~\ref{Fig.5}(e,f). The good agreement between the \emph{ab-initio} EMH and the DFT results demonstrates that even the \(l=1\) truncation is enough for capturing the magnetization-direction dependence of the Hamiltonian with good accuracy. Moreover, this approach can be readily extended to higher orders of \(l\), thereby further improving the accuracy.

\clearpage
\clearpage

\section{Details of first-principles calculations}

The first-principles calculation of monolayer MnBi$_2$Te$_4$ were conducted via the Vienna ab initio simulation package (VASP)~\cite{kresse1996efficient,kresse1996efficiency}. The projector augmented wave (PAW) pseudopotentials~\cite{blochl1994projector,kresse1999ultrasoft} were adopted in the calculation and generalized gradient approximation (GGA)~\cite{perdew1996gga} of the Perdew-Burke-Ernzerhof (PBE) functional~\cite{perdew1998reply} was selected as the exchange correlation potential. The 2D Brillouin zone (BZ) was sampled using Monkhorst-Pack k-mesh~\cite{monkhorst1976special} with a size of $15\times15\times1$. The energy cutoff was set to 600 eV and the energy convergence criteria was $10^{-6}$ eV. A vacuum space ($\sim35$ \AA) is included to avoid interactions between neighboring slabs. Since transition-metal elements have important correlation effects, we adopt the DFT+$U$ method~\cite{dudarev1998electron} with a $U$ value of Mn set at 4 eV. Moreover, we construct the Wannier tight-binding model Hamiltonian with Mn $d$ orbitals, Bi $p$ orbitals, and Te $p$ orbitals, by the maximally localized Wannier functions (MLWF) method via the WANNIER90~\cite{marzari1997maximally,mostofi2008wannier90}.

\bibliography{supref}